\documentclass[a4paper,aps,onecolumn,nofootinbib]{revtex4}
\RequirePackage[colorlinks,hyperindex]{hyperref}
\RequirePackage[english]{babel}
\RequirePackage[latin1]{inputenc}
\RequirePackage[T1]{fontenc}
\RequirePackage{mathrsfs}
\RequirePackage{amsmath}
\RequirePackage{amssymb}
\RequirePackage{amsbsy}
\RequirePackage{color}
\RequirePackage{bm}
\hypersetup{colorlinks=true,breaklinks=true,urlcolor=blue,linkcolor=red}
\begin{document}
\title{\bf{A note on spinor fields in spherical symmetry}}
\author{Stefano Vignolo$^{c}$\!\!\! $^{\hbar}$\!\!\! $^{G}$\footnote{stefano.vignolo@unige.it},
Luca Fabbri$^{c}$\!\!\! $^{\hbar}$\!\!\! $^{G}$\footnote{luca.fabbri@unige.it}}
\affiliation{$^{c}$DIME, Universit\`{a} di Genova, Via all'Opera Pia 15, 16145 Genova, ITALY\\
$^{\hbar}$INFN, Sezione di Genova, Via Dodecaneso 33, 16146 Genova, ITALY\\
$^{G}$GNFM, Istituto Nazionale di Alta Matematica, P.le Aldo Moro 5, 00185 Roma, ITALY}
\date{\today}
\begin{abstract}
By employing the polar re-formulation, we show that there are no solutions of the Dirac equations in spherical symmetry when the spinor is required to satisfy the same symmetries as the space-time via the Lie derivative.
\end{abstract}
\maketitle
\section{Introduction}
When considering physical systems constituted by Dirac spinors, Maxwell electrodynamics, Yang-Mills fields, or Einstein gravity, a situation frequently encountered is that of spherical symmetry in stationary configurations \cite{c-c-d-n, m-r, Saha:2018ufp, Bronnikov:2019nqa, bs-k}.

When Dirac spinors are considered, however, problems may arise, due to the fact that spin can never be reduced to zero and so spherical symmetry may not be possible. In this direction, no-go results were obtained, with spherical symmetry recovered only when the spin was made to vanish, as in spinorial singlets \cite{fss, fss1, fss2, fss3, fss4}.

Whether there are dynamical solutions of the Dirac equations with spherical symmetry in presence of spin is, to our knowledge, still an open problem. In this paper, we will tackle this problem, implementing spherical symmetry by requiring the vanishing of the Lie derivative along the Killing vectors of rotations, for spinor fields written by means of the polar re-formulation \cite{Fabbri:2023onb}. The polar re-formulation of spinor fields consists in having the spinorial components re-arranged in such a way that in their final form they are expressed in terms of the bi-linear spinor quantities. The advantage in this is that, after such a re-formulation, the Dirac theory is written in a form in which all quantities are real and independent on the representation of the Clifford matrices, as well as independent on the choice of frames.

Within this polar re-formulation of spinors, one can see under what conditions a given symmetry implemented via the vanishing of the Lie derivative of the spinor field (strong Lie invariance) and the same symmetry implemented via the vanishing of the Lie derivative of all bi-linear spinor quantities (weak Lie invariance) are equivalent \cite{Fabbri:2023dgv}. In this paper we prove that there do not exist solutions of the Dirac equation which are weakly Lie invariant under spherical symmetry. Since strong Lie invariance implies weak Lie invariance, our outcome holds for strong Lie invariance too. 

The paper is organized as follows: in Sec. \ref{II} we recall the polar form of spinors; in Sec. \ref{III} we implement spherical symmetry for stationary space-times and for spinors in polar form, deriving incompatibilities with the Dirac equation.

Greek/Latin indices indicate coordinate/world indices; we use the so-called ``mostly negative'' signature.
\section{Spinor fields in polar form}\label{II}
We recall the basic conventions and notation with the Clifford matrices $\boldsymbol{\gamma}^{i}$ obeying the usual anti-commutation rules and from which $\boldsymbol{\sigma}_{ik}\!=\![\boldsymbol{\gamma}_{i},\boldsymbol{\gamma}_{k}]/4$ are the generators of the complex Lorentz group. As we will work in $(1\!+\!3)$-dimensional space-times, we have that $2i\boldsymbol{\sigma}_{ab}\!=\!\varepsilon_{abcd}\boldsymbol{\pi}\boldsymbol{\sigma}^{cd}$ implicitly defines the parity-odd matrix $\boldsymbol{\pi}$ (this is usually denoted as a gamma with an index five, but as this is not a true index we prefer to use an index-free notation). Spinor and adjoint spinor are tied by $\overline{\psi}\!=\!\psi^{\dagger}\boldsymbol{\gamma}^{0}$ and with them the bi-linear spinor quantities are defined as
\begin{eqnarray}
&\Sigma^{ab}\!=\!2\overline{\psi}\boldsymbol{\sigma}^{ab}\boldsymbol{\pi}\psi\ \ \ \
\ \ \ \ \ \ \ \ M^{ab}\!=\!2i\overline{\psi}\boldsymbol{\sigma}^{ab}\psi\\
&S^{a}\!=\!\overline{\psi}\boldsymbol{\gamma}^{a}\boldsymbol{\pi}\psi\ \ \ \
\ \ \ \ \ \ \ \ U^{a}\!=\!\overline{\psi}\boldsymbol{\gamma}^{a}\psi\\
&\Theta\!=\!i\overline{\psi}\boldsymbol{\pi}\psi\ \ \ \
\ \ \ \ \ \ \ \ \Phi\!=\!\overline{\psi}\psi,
\end{eqnarray}
all of which being real tensors, with the tensors on the right and the pseudo-tensors on the left column. In the general case in which $\Theta^{2}\!+\!\Phi^{2}\!\neq\!0$, the two tensors can be expressed as
\begin{gather}
\Sigma^{ij}(\Theta^{2}\!+\!\Phi^{2})\!=\!\Phi U^{[i}S^{j]}
\!-\!\varepsilon^{ijab}\Theta U_{a}S_{b}\ \ \ \ \ \ \ \ \ \ \ \ \ \ \ \
M_{ab}(\Theta^{2}\!+\!\Phi^{2})\!=\!\Phi U^{j}S^{k}\varepsilon_{jkab}
\!+\!\Theta U_{[a}S_{b]}
\end{gather}
in terms of the two vectors and the two scalars. These verify $U_{a}U^{a}\!=\!-S_{a}S^{a}\!=\!\Theta^{2}\!+\!\Phi^{2}$ and $U_{a}S^{a}\!=\!0$ known as Fierz identities. When $\Theta^{2}\!+\!\Phi^{2}\!\neq\!0$ spinor fields can be written, in chiral representation, in polar form, as
\begin{eqnarray}
&\psi\!=\!\phi\ e^{-\frac{i}{2}\beta\boldsymbol{\pi}}
\ \boldsymbol{L}^{-1}\left(\begin{tabular}{c}
$1$\\
$0$\\
$1$\\
$0$
\end{tabular}\right)
\label{spinor}
\end{eqnarray}
for a pair of functions $\phi$ and $\beta$ and for some $\boldsymbol{L}$ with the structure of a spinor transformation \cite{jl1}. In polar form
\begin{eqnarray}
&\Theta\!=\!2\phi^{2}\sin{\beta}\ \ \ \ \ \ \ \ \ \ \ \ \ \ \ \ \Phi\!=\!2\phi^{2}\cos{\beta}\label{scalars}
\end{eqnarray}
showing that $\beta$ and $\phi$ are a pseudo-scalar and a scalar, called chiral angle and density, and
\begin{eqnarray}
&S^{a}\!=\!2\phi^{2}s^{a}\ \ \ \ \ \ \ \ \ \ \ \ \ \ \ \ U^{a}\!=\!2\phi^{2}u^{a}\label{vectors}
\end{eqnarray}
as the spin axial-vector and velocity vector. Consequently we have $u_{a}u^{a}\!=\!-s_{a}s^{a}\!=\!1$ and $u_{a}s^{a}\!=\!0$ as reduced Fierz identities. Because the logarithmic derivative of an element of a Lie group belongs to its Lie algebra, we have
\begin{eqnarray}
&\boldsymbol{L}^{-1}\partial_{\mu}\boldsymbol{L}\!=\!iq\partial_{\mu}\zeta\mathbb{I}
\!+\!\frac{1}{2}\partial_{\mu}\zeta_{ij}\boldsymbol{\sigma}^{ij}\label{spintrans}
\end{eqnarray}
for some $\partial_{\mu}\zeta$ and $\partial_{\mu}\zeta_{ij}$ \cite{Fabbri:2021mfc}. Given a gauge potential $A_{\mu}$ and a spin connection $C_{ij\mu}$ we define
\begin{eqnarray}
&\partial_{\mu}\zeta_{ij}\!-\!C_{ij\mu}\!\equiv\!R_{ij\mu}\label{R}\\
&q(\partial_{\mu}\zeta\!-\!A_{\mu})\!\equiv\!P_{\mu}\label{P}
\end{eqnarray}
which are proven to be a real tensor and a gauge-covariant vector called space-time and gauge tensorial connections \cite{Fabbri:2024lyu}. Equipped with these two tensorial connections, we can write
\begin{eqnarray}
&\boldsymbol{\nabla}_{\mu}\psi\!=\!(\nabla_{\mu}\ln{\phi}\mathbb{I}
\!-\!\frac{i}{2}\nabla_{\mu}\beta\boldsymbol{\pi}
\!-\!\frac{1}{2}R_{ij\mu}\boldsymbol{\sigma}^{ij}
\!-\!iP_{\mu}\mathbb{I})\psi
\label{decspinder}
\end{eqnarray}
as the covariant derivative of spinor fields in polar form. Notice that from this expression we can see how the gauge tensorial connection can be identified with the momentum of the particle. We have
\begin{eqnarray}
&\nabla_{\mu}s_{\nu}\!=\!s^{\alpha}R_{\alpha\nu\mu}\ \ \ \ \ \ \ \
\ \ \ \ \ \ \ \ \nabla_{\mu}u_{\nu}\!=\!u^{\alpha}R_{\alpha\nu\mu}\label{ds-du}
\end{eqnarray}
as general identities. The two tensorial connections verify
\begin{eqnarray}
&R_{\alpha\rho\mu\nu}\!=\!-(\nabla_{\mu}R_{\alpha\rho\nu}\!-\!\nabla_{\nu}R_{\alpha\rho\mu}
\!+\!R_{\alpha\kappa\mu}R_{\eta\rho\nu}g^{\kappa\eta}
\!-\!R_{\alpha\kappa\nu}R_{\eta\rho\mu}g^{\kappa\eta})\label{Riemann}\\
&qF_{\mu\nu}\!=\!-(\nabla_{\mu}P_{\nu}\!-\!\nabla_{\nu}P_{\mu})\label{Faraday}
\end{eqnarray}
where $R_{\alpha\rho\mu\nu}$ is the Riemann curvature and $F_{\mu\nu}$ is the Maxwell strength.

The dynamics is established by the Dirac equation
\begin{eqnarray}\label{equazioneDirac_generica}
&i\boldsymbol{\gamma}^{\mu}\boldsymbol{\nabla}_{\mu}\psi\!-\!m\psi\!=\!0:
\end{eqnarray}
in polar form equation (\ref{equazioneDirac_generica}) is equivalently written as
\begin{eqnarray}
&\nabla_{\mu}\ln{\phi^{2}}\!+\!R_{\mu}
\!-\!2P^{\rho}u^{\nu}s^{\alpha}\varepsilon_{\mu\rho\nu\alpha}\!+\!2ms_{\mu}\sin{\beta}\!=\!0
\label{D1}\\
&\nabla_{\mu}\beta\!+\!B_{\mu}
\!-\!2P^{\iota}u_{[\iota}s_{\mu]}\!+\!2ms_{\mu}\cos{\beta}\!=\!0
\label{D2}
\end{eqnarray}
where $R_{\mu}\!=\!R_{\mu\nu}^{\phantom{\mu\nu}\nu}$ and $B_{\mu}\!=\!\frac{1}{2}\varepsilon_{\mu\alpha\nu\iota}R^{\alpha\nu\iota}$ were introduced \cite{Fabbri:2023onb}. We will next proceed to investigate spherical symmetry.
\section{Spherical symmetry and stationary space-times}\label{III}
The spherical symmetry is implemented by requiring the invariance under the group of continuous rotations in the $3$-dimensional space. Therefore we will have to implement the vanishing of the Lie derivative along the Killing vectors, described, in spherical coordinates $(t,r,\theta,\varphi)$, as
\begin{gather}
\xi_{1}\!=\!-\cos{\varphi}\frac{\partial}{\partial \theta}
+\!\sin{\varphi}\cot{\theta}\frac{\partial}{\partial \varphi},\\
\xi_{2}\!=\!\sin{\varphi}\frac{\partial}{\partial \theta}
+\!\cos{\varphi}\cot{\theta}\frac{\partial}{\partial \varphi},\\
\xi_{3}\!=\!\frac{\partial}{\partial \varphi}.
\label{Killing}
\end{gather}
In this note we will also consider stationary cases. We will have the vanishing of the Lie derivative along the Killing vector given by
\begin{eqnarray}
\xi_{0}\!=\!\frac{\partial}{\partial t}.
\end{eqnarray}
When all above conditions are taken into account, one can see that the most general stationary spherically symmetric metric is in the form
\begin{equation}
ds^{2}\!=\!e^{2A}dt \otimes dt+e^{(A+B)}\sinh{\eta}(dt \otimes dr+dr \otimes dt)
-e^{2B}dr \otimes dr-e^{2C}d\theta \otimes d\theta
-e^{2C}(\sin{\theta})^{2}d\varphi \otimes d\varphi
\label{metric}
\end{equation}
with $A\!=\!A(r)$, $B\!=\!B(r)$, $C\!=\!C(r)$ and $\eta\!=\!\eta(r)$ generic functions. This metric generates the connection
\begin{subequations}
\begin{eqnarray}
&\Lambda^{t}_{tt}\!=\!-e^{(A-B)}\frac{A'\tanh{\eta}}{\cosh{\eta}}\ \ \ \ \ \ \ \ \ \ \ \ \ \ \ \
\Lambda^{t}_{tr}\!=\!\frac{A'}{(\cosh{\eta})^{2}}\ \ \ \ \ \ \ \ \ \ \ \ \ \ \ \
\Lambda^{t}_{rr}\!=\!e^{(B-A)}\frac{A'\tanh{\eta}+\eta'}{\cosh{\eta}}\\
&\Lambda^{t}_{\theta\theta}\!=\!e^{(2C-B-A)}\frac{C'\tanh{\eta}}{\cosh{\eta}}\ \ \ \ \ \ \ \
\Lambda^{t}_{\varphi\varphi}\!=\!e^{(2C-B-A)}\frac{C'\tanh{\eta}}{\cosh{\eta}}(\sin{\theta})^{2}\\
&\Lambda^{r}_{tt}\!=\!e^{2(A-B)}\frac{A'}{(\cosh{\eta})^{2}}\ \ \ \ \ \ \ \ \ \ \ \ \ \ \ \
\Lambda^{r}_{rt}\!=\!e^{(A-B)}\frac{A'\tanh{\eta}}{\cosh{\eta}}\ \ \ \ \ \ \ \ \ \ \ \ \ \ \ \
\Lambda^{r}_{rr}\!=\!B'\!+\!A'(\tanh{\eta})^{2}\!+\!\eta'\tanh{\eta}\\
&\Lambda^{r}_{\theta\theta}\!=\!-e^{2(C-B)}\frac{C'}{(\cosh{\eta})^{2}}\ \ \ \ \ \ \ \ \ \ \ \ \ \ \Lambda^{r}_{\varphi\varphi}\!=\!-e^{2(C-B)}\frac{C'}{(\cosh{\eta})^{2}}(\sin{\theta})^{2}\\
&\Lambda^{\theta}_{\theta r}\!=\!\Lambda^{\varphi}_{\varphi r}\!=\!C'\ \ \ \ \ \ \ \
\Lambda^{\theta}_{\varphi\varphi}\!=\!-\cos{\theta}\sin{\theta}\ \ \ \ \ \ \ \
\Lambda^{\varphi}_{\varphi\theta}=\cot{\theta}
\end{eqnarray}\label{conn}
\end{subequations}
and consequently the curvature
\begin{subequations}
\begin{eqnarray}
&R_{t\theta r\theta}(\sin{\theta})^{2}\!=\!R_{t\varphi r\varphi}
\!=\!-e^{(A-B+2C)}\frac{A'C'\tanh{\eta}}{\cosh{\eta}}(\sin{\theta})^{2}\\
&R_{trtr}\!=\!-e^{2A}[A''\!-\!A'(\eta'\tanh{\eta}\!-\!A'\!+\!B')]\\
&R_{t\theta t\theta}(\sin{\theta})^{2}\!=\!R_{t\varphi t\varphi}
\!=\!-e^{(2A-2B+2C)}\frac{A'C'}{(\cosh{\eta})^{2}}(\sin{\theta})^{2}\\
&R_{r\theta r\theta}(\sin{\theta})^{2}\!=\!R_{r\varphi r\varphi}
\!=\!e^{2C}[C''-\!C'(A'(\tanh{\eta})^{2}\!+\!\eta'\tanh{\eta}\!-\!C'\!+\!B')](\sin{\theta})^{2}\\
&R_{\theta\varphi\theta\varphi}\!=\!e^{2(C-B)}\left[\left(\frac{e^{C}C'}{\cosh{\eta}}\right)^{2}
\!-\!e^{2B}\right](\sin{\theta})^{2}
\end{eqnarray}\label{curvature}
\end{subequations}
where it is meant that all non-written components are null.

The same conditions of vanishing of the Lie derivative along all Killing vectors for the bi-linear spinor quantities leads to $\phi\!=\!\phi(r)$ and $\beta\!=\!\beta(r)$ and
\begin{eqnarray}
&s=e^{A}\sinh{(\alpha\!+\!\eta)}dt-e^{B}\cosh{\alpha}dr\ \ \ \ \ \ \ \ \ \ \ \ \ \ \ \ \ \ \ \ \ \ \ \ u=e^{A}\cosh{(\alpha\!+\!\eta)}dt-e^{B}\sinh{\alpha}dr\label{vecs}
\end{eqnarray}
where $\alpha\!=\!\alpha(r)$ is a general function \cite{Fabbri:2023dgv}. Because the structure of these vectors is fixed, relations (\ref{ds-du}) will determine the structure of the space-time tensorial connection. Expanding (\ref{ds-du}) and removing all zero components gives
\begin{subequations}
\begin{eqnarray}
&-s_{t}\Lambda^{t}_{tt}
\!-\!s_{r}\Lambda^{r}_{tt}
\!=\!s^{r}R_{rt t}\ \ \ \ \ \ \ \ \ \ \ \ \ \ \ \
-u_{t}\Lambda^{t}_{tt}
\!-\!u_{r}\Lambda^{r}_{tt}
\!=\!u^{r}R_{rt t}\\
&\partial_{r}s_{t}
\!-\!s_{t}\Lambda^{t}_{tr}
\!-\!s_{r}\Lambda^{r}_{tr}
\!=\!s^{r}R_{rt r}\ \ \ \ \ \ \ \ \ \ \ \ \ \ \ \
\partial_{r}u_{t}
\!-\!u_{t}\Lambda^{t}_{tr}
\!-\!u_{r}\Lambda^{r}_{tr}
\!=\!u^{r}R_{rt r}\\
&0\!=\!R_{rt\theta}\ \ \ \ \ \ \ \ \ \ \ \ \ \ \ \
0\!=\!R_{rt\varphi}\\
&-s_{t}\Lambda^{t}_{tr}
\!-\!s_{r}\Lambda^{r}_{tr}
\!=\!s^{t}R_{tr t}\ \ \ \ \ \ \ \ \ \ \ \ \ \ \ \
-u_{t}\Lambda^{t}_{tr}
\!-\!u_{r}\Lambda^{r}_{tr}
\!=\!u^{t}R_{tr t}\\
&\partial_{r}s_{r}
\!-\!s_{t}\Lambda^{t}_{rr}
\!-\!s_{r}\Lambda^{r}_{rr}
\!=\!s^{t}R_{tr r}\ \ \ \ \ \ \ \ \ \ \ \ \ \ \ \
\partial_{r}u_{r}
\!-\!u_{t}\Lambda^{t}_{rr}
\!-\!u_{r}\Lambda^{r}_{rr}
\!=\!u^{t}R_{tr r}\\
&0\!=\!s^{t}R_{t\theta t}
\!+\!s^{r}R_{r\theta t}\ \ \ \ \ \ \ \ \ \ \ \ \ \ \ \
0\!=\!u^{t}R_{t\theta t}
\!+\!u^{r}R_{r\theta t}\\
&0\!=\!s^{t}R_{t\theta r}
\!+\!s^{r}R_{r\theta r}\ \ \ \ \ \ \ \ \ \ \ \ \ \ \ \
0\!=\!u^{t}R_{t\theta r}
\!+\!u^{r}R_{r\theta r}\\
&-s_{t}\Lambda^{t}_{\theta\theta}
\!-\!s_{r}\Lambda^{r}_{\theta\theta}
\!=\!s^{t}R_{t\theta\theta}
\!+\!s^{r}R_{r\theta\theta}\ \ \ \ \ \ \ \ \ \ \ \ \ \ \ \
-u_{t}\Lambda^{t}_{\theta\theta}
\!-\!u_{r}\Lambda^{r}_{\theta\theta}
\!=\!u^{t}R_{t\theta\theta}
\!+\!u^{r}R_{r\theta\theta}\\
&0\!=\!s^{t}R_{t\theta\varphi}
\!+\!s^{r}R_{r\theta\varphi}\ \ \ \ \ \ \ \ \ \ \ \ \ \ \ \
0\!=\!u^{t}R_{t\theta\varphi}
\!+\!u^{r}R_{r\theta\varphi}\\
&0\!=\!s^{t}R_{t\varphi t}
\!+\!s^{r}R_{r\varphi t}\ \ \ \ \ \ \ \ \ \ \ \ \ \ \ \
0\!=\!u^{t}R_{t\varphi t}
\!+\!u^{r}R_{r\varphi t}\\
&0\!=\!s^{t}R_{t\varphi r}
\!+\!s^{r}R_{r\varphi r}\ \ \ \ \ \ \ \ \ \ \ \ \ \ \ \
0\!=\!u^{t}R_{t\varphi r}
\!+\!u^{r}R_{r\varphi r}\\
&0\!=\!s^{t}R_{t\varphi\theta}
\!+\!s^{r}R_{r\varphi\theta}\ \ \ \ \ \ \ \ \ \ \ \ \ \ \ \
0\!=\!u^{t}R_{t\varphi\theta}
\!+\!u^{r}R_{r\varphi\theta}\\
&-s_{t}\Lambda^{t}_{\varphi\varphi}
\!-\!s_{r}\Lambda^{r}_{\varphi\varphi}
\!=\!s^{t}R_{t\varphi\varphi}
\!+\!s^{r}R_{r\varphi\varphi}\ \ \ \ \ \ \ \ \ \ \ \ \ \ \ \
-u_{t}\Lambda^{t}_{\varphi\varphi}
\!-\!u_{r}\Lambda^{r}_{\varphi\varphi}
\!=\!u^{t}R_{t\varphi\varphi}
\!+\!u^{r}R_{r\varphi\varphi}
\end{eqnarray}\label{expansion}
\end{subequations}
(identical conditions have not been repeated). For all components that are not zero, substituting (\ref{vecs}) explicitly as
\begin{eqnarray}
&s_{t}=e^{A}\sinh{(\alpha\!+\!\eta)}\ \ \ \ \ \ \ \ \ \ \ \ \ \ \ \
u_{t}=e^{A}\cosh{(\alpha\!+\!\eta)}\\
&s_{r}=-e^{B}\cosh{\alpha}\ \ \ \ \ \ \ \ \ \ \ \ \ \ \ \
u_{r}=-e^{B}\sinh{\alpha}
\end{eqnarray}
\begin{eqnarray}
&s^{t}\!=\!e^{-A}\sinh{\alpha}/\cosh{\eta}\ \ \ \ \ \ \ \ \ \ \ \ \ \ \ \
u^{t}\!=\!e^{-A}\cosh{\alpha}/\cosh{\eta}\\
&s^{r}\!=\!e^{-B}\cosh{(\alpha\!+\!\eta)}/\cosh{\eta}\ \ \ \ \ \ \ \ \ \ \ \ \ \ \ \
u^{r}\!=\!e^{-B}\sinh{(\alpha\!+\!\eta)}/\cosh{\eta}
\end{eqnarray}
as well as the connection (\ref{conn}) allows us to reach
\begin{subequations}
\begin{eqnarray}
&e^{2A}A'\!=\!R_{rtt}\ \ \ \ \ \ \ \ \ \ \ \ \ \ \ \ \ \ \ \ \ \ \ \ \ \ \ \ \ \ \ \ \ \ \ \ \ \
-e^{(A+B)}[(\alpha'\!+\!\eta')\cosh{\eta}\!+\!A'\sinh{\eta}]\!=\!R_{trr}\\
&e^{(B-A)}\sinh{\alpha}R_{t\theta t}\!=\!-\cosh{(\alpha\!+\!\eta)}R_{r\theta t}\ \ \ \ \ \ \ \
e^{(B-A)}R_{t\theta t}\!=\!-\frac{\sinh{(\alpha+\eta)}}{\cosh{\alpha}}R_{r\theta t}\\
&e^{(B-A)}\sinh{\alpha}R_{t\theta r}\!=\!-\cosh{(\alpha\!+\!\eta)}R_{r\theta r}\ \ \ \ \ \ \ \
e^{(B-A)}R_{t\theta r}\!=\!-\frac{\sinh{(\alpha+\eta)}}{\cosh{\alpha}}R_{r\theta r}\\
&e^{(B-A)}\sinh{\alpha}R_{t\theta\theta}\!=\!
-\cosh{(\alpha\!+\!\eta)}[R_{r\theta\theta}\!+\!e^{2C}C']\ \ \ \ \ \ \ \ \ \ \ \
e^{(B-A)}R_{t\theta\theta}
\!=\!-\frac{\sinh{(\alpha+\eta)}}{\cosh{\alpha}}[R_{r\theta\theta}\!+\!e^{2C}C']\\
&e^{(B-A)}\sinh{\alpha}R_{t\theta\varphi}\!=\!-\cosh{(\alpha\!+\!\eta)}R_{r\theta\varphi}\ \ \ \
e^{(B-A)}R_{t\theta\varphi}\!=\!-\frac{\sinh{(\alpha+\eta)}}{\cosh{\alpha}}R_{r\theta\varphi}\\
&e^{(B-A)}\sinh{\alpha}R_{t\varphi t}\!=\!-\cosh{(\alpha\!+\!\eta)}R_{r\varphi t}\ \ \ \ \ \ \ \
e^{(B-A)}R_{t\varphi t}\!=\!-\frac{\sinh{(\alpha+\eta)}}{\cosh{\alpha}}R_{r\varphi t}\\
&e^{(B-A)}\sinh{\alpha}R_{t\varphi r}\!=\!-\cosh{(\alpha\!+\!\eta)}R_{r\varphi r}\ \ \ \ \ \ \ \
e^{(B-A)}R_{t\varphi r}\!=\!-\frac{\sinh{(\alpha+\eta)}}{\cosh{\alpha}}R_{r\varphi r}\\
&e^{(B-A)}\sinh{\alpha}R_{t\varphi\theta}\!=\!-\cosh{(\alpha\!+\!\eta)}R_{r\varphi\theta}\ \ \ \
e^{(B-A)}R_{t\varphi\theta}\!=\!-\frac{\sinh{(\alpha+\eta)}}{\cosh{\alpha}}R_{r\varphi\theta}\\
&e^{(B-A)}\sinh{\alpha}R_{t\varphi\varphi}\!=\!-\cosh{(\alpha\!+\!\eta)}[R_{r\varphi\varphi}
\!+\!e^{2C}C'(\sin{\theta})^{2}]\ \ \ \
e^{(B-A)}R_{t\varphi\varphi}\!=\!-\frac{\sinh{(\alpha+\eta)}}{\cosh{\alpha}}[R_{r\varphi\varphi}
\!+\!e^{2C}C'(\sin{\theta})^{2}]
\end{eqnarray}
\end{subequations}
(in which the order of (\ref{expansion}) has been respected, but again we have not repeated identical equations). The first line is
\begin{eqnarray}
&R_{rtt}\!=\!e^{2A}A'\\
&R_{trr}\!=\!-e^{(A+B)}[(\alpha'\!+\!\eta')\cosh{\eta}\!+\!A'\sinh{\eta}]
\end{eqnarray}
while all others have the same structure, from which one can easily deduce that the only solution is given when
\begin{eqnarray}
&R_{r\theta\theta}\!=\!-e^{2C}C'\\
&R_{r\varphi\varphi}\!=\!-e^{2C}C'(\sin{\theta})^{2}
\end{eqnarray}
and all other components zero. Because no angular component of velocity and spin exists, the four components $R_{\theta\varphi\mu}$ have never taken place in this argument. They may be found from the fact that the Riemann curvature computed by means of the space-time tensorial connection (\ref{Riemann}) be exactly the Riemann curvature (\ref{curvature}): for example
\begin{eqnarray}
\nonumber
&e^{2(C-B)}\left[\left(\frac{e^{C}C'}{\cosh{\eta}}\right)^{2}
\!-\!e^{2B}\right](\sin{\theta})^{2}
\!=\!R_{\theta\varphi\theta\varphi}\!=\!-\nabla_{\theta}R_{\theta\varphi\varphi}
\!+\!\nabla_{\varphi}R_{\theta\varphi\theta}
\!-\!R_{\theta\kappa\theta}R_{\eta\varphi\varphi}g^{\kappa\eta}
\!+\!R_{\theta\kappa\varphi}R_{\eta\varphi\theta}g^{\kappa\eta}=\\
&=-\partial_{\theta}R_{\theta\varphi\varphi}
\!+\!R_{\theta\varphi\varphi}\cot{\theta}
\!+\!\partial_{\varphi}R_{\theta\varphi\theta}
\!+\!e^{-2B}\left(\frac{e^{2C}C'}{\cosh{\eta}}\right)^{2}(\sin{\theta})^{2}
\end{eqnarray}
from which
\begin{eqnarray}
\partial_{\theta}\left(\frac{R_{\theta\varphi\varphi}}{\sin{\theta}}
\!+\!e^{2C}\cos{\theta}\right)
\!-\!\partial_{\varphi}\left(\frac{R_{\theta\varphi\theta}}{\sin{\theta}}\right)\!=\!0:
\end{eqnarray}
this integrability condition admits the general solution
\begin{gather}
R_{\theta\varphi\varphi}\!=\!e^{2C}\sin{\theta}(\partial_{\varphi}F\!-\!\cos{\theta})\\
R_{\theta\varphi\theta}\!=\!e^{2C}\sin{\theta}\partial_{\theta}F
\end{gather}
where $F\!=\!F(t,r,\theta,\varphi)$ is a generic integration function. Other components of the Riemann curvature would lead to
\begin{eqnarray}
R_{\theta\varphi t}\!=\!e^{2C}\sin{\theta}\partial_{t}F\\
R_{\theta\varphi r}\!=\!e^{2C}\sin{\theta}\partial_{r}F
\end{eqnarray}
and therefore, all in all, we end up with
\begin{subequations}
\begin{gather}
R_{rtt}\!=\!e^{2A}A'\ \ \ \ \ \ \ \ \ \ \ \ \ \ \ \
R_{\theta\varphi t}\!=\!e^{2C}\sin{\theta}\partial_{t}F\\
R_{trr}\!=\!-e^{(A+B)}[(\alpha'\!+\!\eta')\cosh{\eta}\!+\!A'\sinh{\eta}]\ \ \ \ \ \ \ \ \ \ \ \ R_{\theta\varphi r}\!=\!e^{2C}\sin{\theta}\partial_{r}F\\
R_{r\theta\theta}\!=\!-e^{2C}C'\ \ \ \ \ \ \ \ \ \ \ \ \ \ \ \
R_{\theta\varphi\theta}\!=\!e^{2C}\sin{\theta}\partial_{\theta}F\\
R_{r\varphi\varphi}\!=\!-e^{2C}C'(\sin{\theta})^{2}\ \ \ \ \ \ \ \ \ \ \ \ \ \ \ \
R_{\theta\varphi\varphi}\!=\!e^{2C}\sin{\theta}(\partial_{\varphi}F\!-\!\cos{\theta})
\end{gather}\label{TC}
\end{subequations}
as the list of all non-zero components of the space-time tensorial connection.

From these, we can compute the relevant vectors to be
\begin{eqnarray}
&R_{t}\!=\!e^{(A-B)}\frac{\alpha'+\eta'}{\cosh{\eta}}\ \ \ \ \ \ \ \
R_{r}\!=\!A'\!+\!(\alpha'\!+\!\eta')\tanh{\eta}\!+\!2C'\ \ \ \ \ \ \ \
R_{\theta}\!=\!(\cos{\theta}\!-\!\partial_{\varphi}F)/\sin{\theta}\ \ \ \ \ \ \ \
R_{\varphi}\!=\!\sin{\theta}\partial_{\theta}F
\end{eqnarray}
and
\begin{eqnarray}
&B_{t}\!=\!-e^{(A-B)}\frac{1}{\cosh{\eta}}\partial_{r}F\!+\!\tanh{\eta}\partial_{t}F\ \ \ \ \ \ \ \
\ \ \ \ \ \ \ \ B_{r}\!=\!-\tanh{\eta}\partial_{r}F\!-\!e^{(B-A)}\frac{1}{\cosh{\eta}}\partial_{t}F
\end{eqnarray}
used to write the Dirac equations in polar form. Of all $8$ equations, and specifically of the $4$ angular components, two turn out to be identically verified, while the other two are given by
\begin{eqnarray}
&P_{\theta}\!-\!\frac{1}{2}\partial_{\theta}F\!=\!0\\
&P_{\varphi}\!-\!\frac{1}{2}\partial_{\varphi}F\!=\!-\frac{1}{2}\cos{\theta}:
\end{eqnarray}
the last component is responsible for the odd behaviour of the spinor under spherical symmetry. To see that, recall that even in presence of electrodynamics, spherical symmetry requires the two angular components of the potential $A_{\mu}$ to vanish. Hence, with or without electrodynamics, the two angular components of the momentum (\ref{P}) are
\begin{eqnarray}
&P_{\theta}\!\equiv\!q\partial_{\theta}\zeta\\
&P_{\varphi}\!\equiv\!q\partial_{\varphi}\zeta
\end{eqnarray}
and so the last two Dirac equations read
\begin{eqnarray}
&\partial_{\varphi}(q\zeta\!-\!F/2)\!=\!-\frac{1}{2}\cos{\theta}\ \ \ \ \ \ \ \
\ \ \ \ \ \ \ \ \partial_{\theta}(q\zeta\!-\!F/2)\!=\!0
\end{eqnarray}
with the consequence that
\begin{eqnarray}
&\frac{1}{2}\sin{\theta}\!=\!\partial_{\theta}\partial_{\varphi}(q\zeta\!-\!F/2)
\!=\!\partial_{\varphi}\partial_{\theta}(q\zeta\!-\!F/2)\!=\!0
\end{eqnarray}
obtaining a contradiction. This incompatibility is the result of spherical symmetry implemented at the weak level: because weak Lie invariance is implied by strong Lie invariance, incompatibility occurs also at the strong level.

Notice that if we insisted on spherical symmetry on the momentum itself, the last Dirac equation would have been $\cos{\theta}\!=\!0$ and the contradiction would have been obtained with just one Dirac equation. It is also important to remark that, because we have used a general metric, but no dynamical equations for the curvature of the space-time, our result remains valid whether in Einstein gravity or in any of its higher-order derivative extensions.

The obtained incompatibility is the result of spherical symmetry within the Dirac equation. Nevertheless, one may ask whether it is possible to get such incompatibility even with no Dirac equation, at a kinematic level. The answer is affirmative \emph{if} one invokes invariance under parity. Invariance under parity is given by $s_{\nu}'\!=\!s_{\nu}$ and $u_{\nu}'\!=\!u_{\nu}$ for the transformation $\theta'\!=\!\pi-\theta$, which in particular implies $\partial\theta'/\partial\theta\!=\!-1$ and a negative Jacobian: because no vector has $\theta$ components and in the non-zero components nothing depends on $\theta$, parity acts only via the Jacobian, according to $s_{t}'\!=\!-s_{t}$ and $s_{r}'\!=\!-s_{r}$, or $s_{\nu}'\!=\!-s_{\nu}$, with $u_{\nu}$ left unchanged. So, invariance under parity implies $s_{\nu}\!=\!s_{\nu}'\!=\!-s_{\nu}$, and thus $s_{\nu}\!=\!0$: then $s^{a}s_{a}\!=\!0$ and this is incompatible with the fact that $s^{a}s_{a}\!=\!-1$, generating a contradiction \cite{Fabbri:2023dgv}.
\section{Conclusion}
In this work, we have considered a Dirac spinor in stationary spherically symmetric space-times, and we have seen that, when such symmetry is implemented also on the bi-linear spinor quantities, then incompatibilities are met. More precisely, we have shown that there are no solutions of the Dirac equations in spherical symmetry when the spinor is required to satisfy the same symmetries as the space-time via the Lie derivative.
\vspace{10pt}

\textbf{Funding}. This work is done in the framework of the INFN Research Project QGSKY and funded by Next Generation EU via the project ``Geometrical and Topological effects on Quantum Matter (GeTOnQuaM)''.

\

\textbf{Data Availability}. The manuscript does not have associated data in any repository.

\

\textbf{Conflict of interest}. There is no conflict of interest.

\end{document}